\documentclass[twocolumn,aps]{revtex4}

\usepackage{graphicx}

\begin{document}

\preprint{}

\title{Resonator-Aided Single-Atom Detection on a Microfabricated Chip}

\author{Igor Teper, Yu-Ju Lin, and Vladan Vuleti\'{c}}

\affiliation {MIT-Harvard Center for Ultracold Atoms, MIT,
Cambridge, MA 02139.}

\date{\today}

\begin{abstract}
We use an optical cavity to detect single atoms magnetically trapped
on an atom chip. We implement the detection using both fluorescence
into the cavity and reduction in cavity transmission due to the
presence of atoms. In fluorescence, we register 2.0(2) photon counts
per atom, which allows us to detect single atoms with 75$\%$
efficiency in 250 $\mu$s. In absorption, we measure transmission
attenuation of 3.3(3)$\%$ per atom, which allows us to count small
numbers of atoms with a resolution of about 1 atom.
\end{abstract}

\maketitle
    In the past several years, there have been many promising developments
in the field of microfabricated magnetic traps (microtraps) for
ultracold atoms, including experimental realizations of
microtrap-based atom interferometers \cite{Wang05, Schumm05}, atomic
clocks \cite{Treutlein04}, and Bragg reflectors \cite{Gunther05}.
Compared to optical traps, where significant progress in
interferometry \cite{Shin04}, Josephson junctions \cite{Albiez05}
and one-dimensional physics \cite{Tolra04, Paredes04, Kinoshita05}
has been made, microchips offer smaller length scales and tighter
confinement for single traps, which, however, may require working
with small atom numbers. Furthermore, there are many proposed atom
chip experiments, such as the implementation of a Tonks-Girardeau
gas \cite{Olshanii98, Dunjko01} or an atomic Fabry-Perot
interferometer \cite{Carusotto01} in a magnetic trap, that may
greatly benefit from measuring atom statistics and correlations at
the single-atom level, as has recently been demonstrated in a
free-space experiment \cite{Ottl05}. In addition, the preparation
and detection of single atoms in microtraps constitute an important
step toward quantum information processing with neutral atoms, which
could take advantage of the tight, complex, precisely controlled,
and scalable magnetic traps available on microchips
\cite{Calarco00}. In this context, the problem arises of how to
detect small atom numbers in magnetic microtraps close to a
substrate surface with a good signal-to-noise ratio \cite{Horak03,
Long03}.

    While cavity QED experiments \cite{Munstermann99,McKeever04}
can detect and count single atoms in the strong coupling regime,
integrating very-high-finesse Fabry-Perot cavities with atom chips
may prove difficult \cite{Long03}. Observing small atom numbers
through fluorescence in a magneto-optical trap (MOT) \cite{Chuu05}
or in MOT-loaded dipole traps \cite{Schlosser01,Kuhr01} requires
long measurement times, and is also not easily compatible with chip
traps. While experimental progress has recently been made in
incorporating fiber resonators \cite{Quinto-Su04} and microcavities
\cite{Trupke05} into atom chips, the capabilities of such detection
methods remain to be established. Recently, a low-finesse concentric
cavity was used for sensitive detection of atoms in a macroscopic
magnetic waveguide in a free-space geometry \cite{Haase06}.

    Atom detection can be implemented via fluorescence
\cite{Chuu05,Schlosser01,Kuhr01} or absorption methods
\cite{Ottl05,Munstermann99,McKeever04}. To compare the two methods,
consider a sample of $a$ atoms, where each atom on average scatters
$m$ photons, and the imaging system detects a fraction $\alpha$ of
the photons. If background counts can be neglected, the atom number
uncertainty $\Delta a$ in fluorescence detection is given by the
ratio between the photon shot noise, $\sqrt{a \alpha m}$, and the
signal per atom, $\alpha m$, and equal to $\Delta a=\sqrt{a/\alpha
m}$. Consequently, in fluorescence measurements, the resolution
decreases as the square root of the atom number.

     For the equivalent absorption measurement, we assume the same number
$m$ of scattered photons per atom, an absorption beam matched to the
collection optics, and low total absorption. Then, the number of
incoming photons is $m/\alpha$ with a shot noise $\sqrt{m/\alpha}$.
This results in an atom number uncertainty of $\Delta
a=\sqrt{1/\alpha m}$, independent of atom number and equal to the
fluorescence resolution for $a=1$. Thus, while fluorescence
detection may provide a superior single-atom detector if background
counts are negligible, absorption detection should perform better as
a single-shot atom counter. Cavity-aided detection is attractive for
both fluorescence and absorption methods, since the emission of
light into the cavity is enhanced by the Purcell factor $F/\pi$,
where $F$ is the cavity finesse \cite{Purcell46}.

\begin{figure}
\includegraphics[width=3.3in]{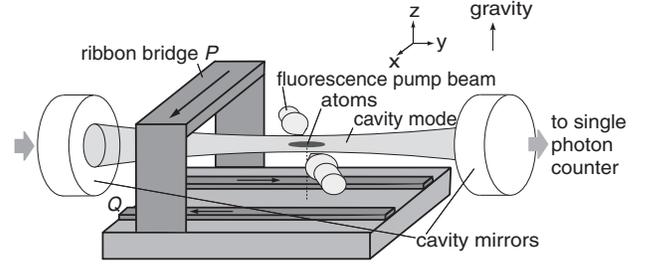}
\caption{Cavity and microfabricated chip (not to scale). The chip
wires ($Q$) generate a 2D quadrupole field in the $xz$ plane. The
ribbon ($P$) in combination with an external field gradient creates
the confinement along $y$. The atoms are probed either with a pump
beam from the side to induce fluorescence into the cavity or through
the absorption of a beam coupled through the cavity.}\label{fig1}
\end{figure}

    In this paper, we investigate the detection and counting of small
numbers of atoms in a magnetic microtrap using a macroscopic,
medium-finesse Fabry-Perot cavity employing both fluorescence and
absorption detection techniques. Using shot-noise-limited atom
preparation down to 1 atom, we achieve single-atom sensitivity in
the fluorescence scheme, and a resolution of about 1 atom in
absorption.

    Our experimental setup (Fig.~\ref{fig1}) is similar to that described
in Ref. \cite{Lin04}. $^{87}$Rb atoms are trapped and cooled in a
free-space magneto-optical trap, transferred to a magnetic trap,
which is then moved close to the chip and adiabatically transformed
into a Ioffe-Pritchard microtrap. The radial $(xz)$ confinement of
the chip trap is provided by two 2-$\mu$m-high, 500-$\mu$m-wide gold
wires $Q$, whose centers are separated by 1 mm, carrying
antiparallel currents along the $y$ direction, in superposition with
a bias field along $z$. The axial ($y$) confinement is created by a
current through a gold ribbon bridge $P$ along $x$, 500 $\mu$m away
from the chip's surface, in combination with an external field
gradient along $y$. We have mounted high-reflectivity, low-loss
mirrors on opposite sides of the chip to form a 2.66 cm long
near-confocal cavity with TEM$_{00}$ mode waist $w$ of 56 $\mu$m,
finesse $F=8600$, linewidth $\kappa/2\pi=650$ kHz, free spectral
range of 5630 MHz, and transverse mode spacing of 230 MHz, aligned
along the axis of our magnetic trap and located 200 $\mu$m away from
the chip surface, so that the cavity mode passes between the bridge
bond and the chip. The mode waist is displaced longitudinally by 5.6
mm relative to the microtrap. For the fluorescence measurement, a
retroreflected pump beam with a waist of 250 $\mu$m at the position
of the atoms is coupled from the side of the cavity at an angle of
70$^\circ$ to the cavity axis.

    We prepare our atoms by initially loading them into a
Ioffe-Pritchard microtrap located 200 $\mu$m away from the surface
and displaced to the side of the cavity axis in order to prevent
them from being heated by the cavity length stabilization light.  We
then use a fast radiofrequency (RF) evaporation to remove the vast
majority of the atoms, leaving us with a small number of cold atoms
at a typical temperature of 15 $\mu$K. Once we have the sample
ready, we ramp the magnetic field to move the trap into the cavity
mode, turn off the locking light, and perform the fluorescence or
absorption measurement. When located in the cavity, the magnetic
trap has a radial gradient of 200 G/cm, which corresponds to
transverse vibration frequencies around 300 Hz for typical offset
fields, and an axial vibration frequency of 50 Hz.

    Using optics outside the vacuum chamber, we couple
up to 95$\%$ of the light entering the cavity into the TEM$_{00}$
mode. The light exiting the cavity is mode-matched to a single-mode
fiber, which allows for excellent spatial filtering of background
light, and delivered to a single-photon-counting module (SPCM). We
stabilize our cavity to an off-resonant laser of a linewidth
slightly broader than that of the cavity using a Pound-Drever-Hall
scheme. The stability of our cavity lock is 140 kHz/$\sqrt{10
\text{kHz}}$ for a locked cavity, and 400 kHz/$\sqrt{10 \text{kHz}}$
in the several ms after the locking light is extinguished, which is
when we perform our measurements.

    Both absorption and fluorescence signals in cavity-aided detection
depend on the atoms' scattering rate of photons into the cavity. For
an atom on the cavity axis, the ratio $\eta$ of its scattering rate
into each of the two directions of the cavity, $\Gamma_{c}$, to its
scattering rate into free space, $\Gamma_{s}$, is given by the
single-atom cooperativity, $\eta=6F/(\pi(wk)^{2})$, where $w$ is the
cavity mode waist size and $k=2\pi/\lambda$ with $\lambda$ as the
optical wavelength resonant with the cavity mode; for our cavity,
$\eta=0.07$. To confirm our atom-cavity coupling experimentally, we
measure the tuning of the transmission resonance by samples large
enough that the atom number can be determined by standard absorption
techniques. The tuning of the cavity resonance by an atom
well-localized at the cavity waist is given by
$\delta\nu=(\kappa/2)(\Gamma/\Delta)N\eta$, where $\Gamma$ is the
linewidth of the atomic transition, $\Delta \gg \Gamma$ is the
detuning between the laser and the atomic transition, and $N$ is the
atom number. The measurement of $\delta\nu$ yields a value of $\eta$
between 0.015 and 0.025, in good agreement with the value we would
expect given our independent measurement of the cloud size, which
reduces $\eta$ compared to the on-axis case.

\begin{figure}
\includegraphics[width=3.3in]{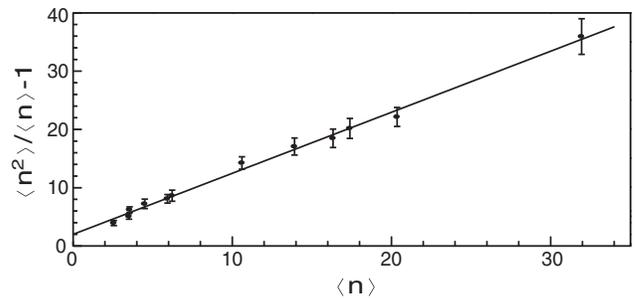}
\caption{Characterization of the atom number preparation noise and
detected photon number per atom for fluorescence detection. $n$
denotes the number of signal photons detected. The atom-atom
correlation function, $g_{aa}$, is given by the slope, and the
average number of photon counts per atom, $\langle p\rangle$, is
given by the y-axis intercept.}\label{fig2}
\end{figure}

    To characterize both the atom number preparation and the number
of photon counts per atom in fluorescence detection,  we illuminate
the atoms with a near-resonant pump beam just above saturation and
count the photons emerging from the cavity within 750 $\mu$s. We
compile histograms of counts for different RF final settings, i.e.,
different average numbers of prepared atoms. Given Poisson
statistics for the detected photons, the following relation can be
derived: $\langle n^2\rangle /\langle n\rangle-1 =g_{aa}\langle
n\rangle+\langle p \rangle$, where $\langle a \rangle$ is the mean
atom number, $\langle p \rangle$ is the mean number of photon counts
per atom, $n=ap$ is the number of signal photon counts, and $g_{aa}$
is the atom-atom correlation function, which should be equal to 1 if
the atoms strictly obey Poisson statistics and equal to $(1+f^{2})$
in the presence of (technical) fractional atom number noise of
magnitude $f$. The values of $\langle n^2\rangle /\langle
n\rangle-1$ can be computed from each histogram independently
without any knowledge about $\langle a\rangle $ or $\langle p\rangle
$, given that we can measure the background count rate
independently, and assuming that fluctuations in the background are
uncorrelated with fluctuations in the signal. The results, along
with a linear fit, are plotted in Fig.~\ref{fig2}. The fit gives a
slope of $g_{aa}=1.05(2)$, which implies that the fractional noise
on our signal is 0.25(10), and therefore Poissonian fluctuations
dominate for the atom numbers we measure, and an intercept of
$\langle p\rangle =1.9(3)$.

\begin{figure}
\includegraphics[width=3.3in]{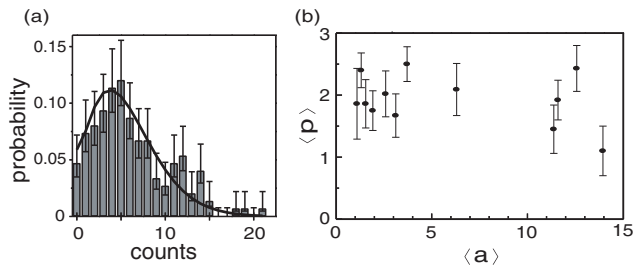}
\caption{(a) Typical normalized histogram of 150 fluorescence
measurements, with Poisson fit to $\langle a \rangle$, the mean
number of atoms, and $\langle p \rangle$, the mean number of photon
counts per atom. In this case, $\langle a \rangle =3.1(4)$ and
$\langle p \rangle=1.7(3)$. Error bars are Poissonian uncertainties.
(b) Results of Poisson fits to 12 different histograms, with error
bars corresponding to 1 standard deviation in $\langle p
\rangle$.}\label{fig3}
\end{figure}

    Having confirmed the Poisson statistics of our atom number preparation,
we can fit the $\langle a \rangle$ and $\langle p \rangle$ for each
histogram individually. A typical histogram with fit and a plot of
the combined results of all histogram fits are shown in
Fig.~\ref{fig3}. To a good approximation, the average number of
photon counts per atom is independent of atom number at $\langle p
\rangle=2.0(2)$ counts/atom, with 0.3 background counts. An average
of the signal time traces yields a 1/$e$ time of $\tau=150$ $\mu$s.

    Since our cavity resonance is much narrower than the atomic line,
the cavity collects predominantly the coherently scattered photons
and only a small fraction of the Mollow triplet \cite{Mollow69}. The
number of photon counts we would expect to detect per atom is thus
given by $\langle p
\rangle=\Gamma_{coh}\tau\eta(\kappa/\gamma)\times CG\times g \times
f\times q\times l$, where $\Gamma_{coh}=\Gamma/8=2\pi\times 760$ kHz
is the maximum coherent scattering rate for the transition,
$\eta=0.07$, $\gamma=2\pi\times 1$ MHz is the linewidth of the
cavity transmission, which is a convolution of the cavity and laser
linewidths, $CG=0.3$ is the averaged Clebsch-Gordan coefficient for
$\sigma^{+}$ intracavity light coming from the scattering process
(the other polarizations are not resonant with the cavity), $g=0.6$
accounts for the finite size of the atomic cloud, $f=0.7$ is the
coupling efficiency into the single-mode fiber, $q=0.58$ is the
quantum efficiency of the SPCM, and $l=0.7$ is the signal reduction
due to measured meachanical cavity vibrations. The combination of
the above factors predicts $\langle p \rangle=1.7$, close to our
measured value.

    In order to quantify the performance of our fluorescence measurement
as a single-atom detector, we reduced the measurement window to 250
$\mu$s, and took a histogram with, on average, less than one atom
prepared. The Poisson fit to the resulting histogram gives $\langle
a \rangle=0.85(8)$ and $\langle p \rangle=1.4(1)$. Combined with a
measured background of 0.07 counts, this means that, if we set our
detection threshold to $\ge$ 1 count, our single-atom detection is
characterized by an atom quantum efficiency of 75$\%$ and a false
detection rate of 7$\%$, at a maximum single-atom count rate of 4
kHz.

\begin{figure}
\includegraphics[width=3.3in]{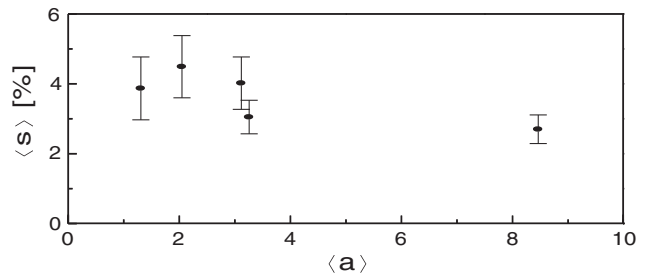}
\caption{Results of Poisson fits to $\langle s \rangle$, the mean
single-atom absorption, and $\langle a \rangle$, the mean atom
number, for 5 different histograms.} \label{fig4}
\end{figure}

    While the fluorescence measurement makes a good single-atom detector,
we expect an absorption measurement to provide better atom number
resolution for $a>1$. For absorption detection, we couple the probe
laser beam into the cavity TEM$_{00}$ mode and monitor the resonant
transmission through the cavity in the presence of atoms. The laser
linewidth is broadened by frequency modulation to 30 MHz, much wider
than $\kappa$, so that the intensity noise on the cavity
transmission due to cavity vibration is negligible compared to the
photon shot noise. The laser is tuned to atomic resonance with an
intracavity saturation parameter equal to 0.2. The expected
transmission reduction due to one atom of a laser resonant with the
atomic transition is $2\eta$. Similarly to fluorescence detection,
we compile histograms for different sets of atom preparation
parameters and fit them, assuming Poisson statistics for both the
atoms and the photons, to determine the mean absorption per atom,
$\langle s \rangle$, and the mean atom number, $\langle a \rangle$.
(A correlation function fit similar to the one for fluorescence
confirms that our atom preparation for absorption has Poisson
statistics, with $g_{aa}=0.94(4)$.) The fitting results for $\langle
s \rangle$ with varying atom number are shown in Fig.~\ref{fig4}.
From these measurements, we obtain $\langle s \rangle=3.3(3)\%$ for
$1<\langle a \rangle<10$, in good agreement with the expected
absorption per atom, $\langle s \rangle =3.2(7)\%$, when geometric
factors due to finite cloud size are taken into account.

\begin{figure}
\includegraphics[width=3.3in]{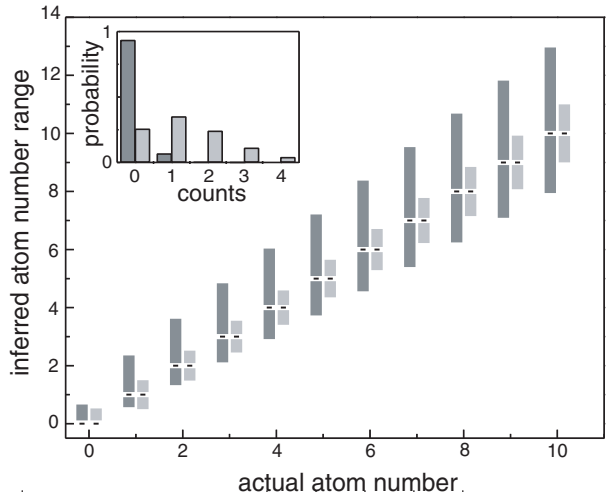}
\caption{Atom number measurement 1-$\sigma$ confidence intervals in
a single shot for fluorescence ($\tau=750$ $\mu$s, dark grey) and
absorption ($\tau=1$ ms, light grey) measurements. The inset shows a
computed normalized photon count distribution due to background
counts (dark grey) and due to photons collected from one atom (light
grey) for fluorescence single-atom detection ($\tau=750$
$\mu$s).}\label{fig5}
\end{figure}

    Using the measured values of $\langle p\rangle =2.0(2)$ counts/atom
for fluorescence detection and $\langle s \rangle=3.3(2)\%$ for
absorption, we can evaluate how well these two methods can determine
the atom number in a single measurement. The expected atom number
uncertainty $\delta a$ using fluorescence (absorption) detection due
to both photon shot noise and the statistical uncertainty in the
mean number of photons per atom, $\langle p\rangle$ (uncertainty in
the mean absorption per atom, $\langle s \rangle$), as well as the
background photon counts (for fluorescence only), is plotted as a
function of atom number in Fig.~\ref{fig5}; the figure also includes
a computed normalized histogram that characterizes the single-atom
detection capability of our fluorescence measurement. For
fluorescence, the atom number resolution is limited by the shot
noise of the collected signal photons, which grows with atom number,
while, for absorption, where the number of collected photons
actually decreases with atom number, the resolution remains nearly
flat, at around 1 atom.

    The demonstrated excellent atom number resolution could be
useful in a variety of microchip experiments. For instance, a
Tonks-Girardeau (TG) gas could be created close to the chip, where
high radial vibration frequencies can be achieved. The sample could
then be moved into the cavity to measure both the atom number and
the density distribution of the gas. As an example, a quantum
degenerate gas of 50 $^{87}$Rb atoms confined in a magnetic trap
with a radial trapping frequency of 20 kHz and an axial frequency of
0.5 Hz would be deep in the TG regime, with $\gamma=
2/(n|a_{\text{1D}}|)$ of 10, where $\gamma \gg 1$ means that the gas
experiences strong fermionization, for a peak one-dimensional number
density $n$ and an effective one-dimensional scattering length
$a_{\text{1D}}$ \cite{Dunjko01}. Then the spatial (30 $\mu$m) and
atom number resolution of our detector would allow one to
distinguish the length and density distribution of this TG gas
($l=300$ $\mu$m) from a corresponding non-fermionized Thomas-Fermi
gas with the same atom number ($l=420$ $\mu$m).

    In conclusion, we have demonstrated $in~situ$ detection of
magnetically trapped atoms on a chip with the aid of a
medium-finesse macroscopic cavity, and characterized the performance
for single-atom detection and for atom number measurements using
both fluorescence and absorption methods. We believe that, due to
their combination of versatility, performance, and ease of use, such
cavity-aided detection schemes can play an important role in a broad
range of applications in integrated atom optics on chips.

    This work was supported by the NSF and DARPA.

\end{document}